\documentclass[conference, final]{IEEEtran}
\IEEEoverridecommandlockouts
\usepackage{amsmath,amssymb,amsfonts}
\usepackage{algorithmic}
\usepackage{graphicx}
\usepackage{textcomp}
\usepackage{xcolor}
\usepackage{bm}
\usepackage{subfigure}
\usepackage{hyperref}
\newcommand{\angstrom}{\mbox{\normalfont\AA}}

\def\BibTeX{{\rm B\kern-.05em{\sc i\kern-.025em b}\kern-.08em
    T\kern-.1667em\lower.7ex\hbox{E}\kern-.125emX}}
\begin{document}

\title{Context-Aware Neural Video Compression on Solar Dynamics Observatory\thanks{ This research is based upon work supported by the National Aeronautics and Space Administration
(NASA), via award number 80NSSC21M0322 under the title
of \emph{Adaptive and Scalable Data Compression for Deep Space
Data Transfer Applications using Deep Learning}.}
}

\author{
     Atefeh Khoshkhahtinat$^\dag$, Ali Zafari$^\dag$, Piyush M. Mehta$^\ddag$, Nasser M. Nasrabadi$^\dag$, Barbara J. Thompson$^\S$,\\ Michael S. F. Kirk$^\S$, Daniel da Silva$^\S$\\
    $^\dag$Dept. of Computer Science \& Electrical Engineering, West Virginia University, WV USA\\
    $^\ddag$Dept. of Mechanical \& Aerospace Engineering, West Virginia University, WV USA\\
    $^\S$NASA Goddard Space Flight Center, MD USA\\
    {
    \tt\small \{\href{mailto:ak00043@mix.wvu.edu}{ak00043},\href{mailto:az00004@mix.wvu.edu}{az00004}\}@mix.wvu.edu,\{\href{mailto:piyush.mehta@mail.wvu.edu}{piyush.mehta},\href{mailto:nasser.nasrabadi@mail.wvu.edu}{nasser.nasrabadi}\}@mail.wvu.edu
    % \tt\small \{\href{mailto:barbara.j.thompson@nasa.gov}{barbara.j.thompson}, \href{mailto:daniel.e.dasilva@nasa.gov}{daniel.e.dasilva}, \href{mailto:michael.s.kirk@nasa.gov}{michael.s.kirk}\}@nasa.gov
    % }
    }\\
    {
    \tt\small \{\href{mailto:barbara.j.thompson@nasa.gov}{barbara.j.thompson},\href{mailto:michael.s.kirk@nasa.gov}{michael.s.kirk},\href{mailto:daniel.e.dasilva@nasa.gov}{daniel.e.dasilva}\}@nasa.gov
    
    }
}

% \author{\IEEEauthorblockN{Ali Zafari}
% \IEEEauthorblockA{\textit{Dept. of Computer Science \& Electrical}\\
% \textit{Engineering, West Virginia University}\\
% Morgantown, WV USA \\
% \href{mailto:az00004@mix.wvu.edu}{az00004@mix.wvu.edu}}
% \and
% \IEEEauthorblockN{Atefeh Khoshkhahtinat}
% \IEEEauthorblockA{\textit{Dept. of Computer Science \& Electrical}\\
% \textit{Engineering, West Virginia University}\\
% Morgantown, WV USA \\
% \href{mailto:ak00043@mix.wvu.edu}{ak00043@mix.wvu.edu}}
% \and
% \IEEEauthorblockN{Piyush M. Mehta}
% \IEEEauthorblockA{\textit{Dept. of Mechanical and Aerospace} \\
% \textit{Engineering, West Virginia University}\\
% Morgantown, WV USA \\
% \href{mailto:piyush.mehta@mail.wvu.edu}{piyush.mehta@mail.wvu.edu}}
% \and
% \IEEEauthorblockN{Nasser M. Nasrabadi}
% \IEEEauthorblockA{\textit{Dept. of Computer Science \& Electrical}\\
% \textit{Engineering, West Virginia University}\\
% Morgantown, WV USA \\
% \href{mailto:nasser.nasrabadi@mail.wvu.edu}{nasser.nasrabadi@mail.wvu.edu}}
% \and
% \IEEEauthorblockN{Barbara J. Thompson}
% \IEEEauthorblockA{\textit{NASA Goddard Space Flight Center}\\
% Greenbelt, MD USA \\
% \href{mailto:barbara.j.thompson@nasa.gov}{barbara.j.thompson@nasa.gov}}
% \and
% \IEEEauthorblockN{Daniel da Silva}
% \IEEEauthorblockA{\textit{NASA Goddard Space Flight Center}\\
% Greenbelt, MD USA \\
% \href{mailto:daniel.e.dasilva@nasa.gov}{daniel.e.dasilva@nasa.gov}}
% \and
% \IEEEauthorblockN{Michael S. F. Kirk}
% \IEEEauthorblockA{\centerline{\textit{NASA Goddard Space Flight Center}}\\
% Greenbelt, MD USA \\
% \href{mailto:michael.s.kirk@nasa.gov}{michael.s.kirk@nasa.gov}}
% }

\maketitle

\begin{abstract}

%Solar Dynamics Observatory (SDO) mission of NASA gathers 1.4 Tera-bytes of data each day orbiting in space. To transmit this huge amount of data to be used in downstream scientific research, it is required to compress the raw data efficiently. This data includes images of the Sun captured at different wavelengths.

NASA’s Solar Dynamics Observatory (SDO) mission collects large data volumes of the Sun’s daily activity. Data compression is crucial for space missions to reduce data storage and video bandwidth requirements by eliminating redundancies in the data. In this paper, we present a novel neural Transformer-based video compression approach specifically designed for the SDO images. Our primary objective is to efficiently exploit the temporal and spatial redundancies inherent in solar images to obtain a high compression ratio. Our proposed architecture benefits from a novel Transformer block called \textit{Fused Local-aware Window (FLaWin)}, which incorporates window-based self-attention modules and an efficient \textit{fused local-aware feed-forward (FLaFF)} network. This architectural design allows us to simultaneously capture short-range and long-range information while facilitating the extraction of rich and diverse contextual representations. Moreover, this design choice results in reduced computational complexity. Experimental results demonstrate the significant contribution of the FLaWin Transformer block to the compression performance, outperforming conventional hand-engineered video codecs such as H.264 and H.265 in terms of rate-distortion trade-off. 
\end{abstract}


\section{\textbf{Introduction}}

NASA’s Solar Dynamics Observatory (SDO) mission gathers 1.4 terabytes of data that can be used to understand the effect of the Sun on the Earth each day \cite{schou2012hmi}. Due to the problem of onboard data storage and bandwidth limitations, data compression is inevitable in space missions. Both hand-crafted \cite{fischer2017jpeg2000eve} and neural-based \cite{zafari2022attention,zafari2022neural} codecs have been proposed to tackle the challenge of data compression on this space mission.

Recently, neural image/video compression methods have achieved remarkable performance compared with their traditional counterparts \cite{yang2022introduction}. All the compression methods attempt to exploit the redundancies in images and videos. There are three types of redundancies in image signals: spatial redundancy, visual redundancy, and statistical redundancy. In addition to the above-mentioned redundancies in image signals, video signals inherit the advantage of temporal redundancy, which allows video compression to obtain a higher compression ratio compared with the still image compression \cite{ma2019image}.

In image/video compression, a transformation function is utilized to map the data to an uncorrelated latent space. The more decorrelated and energy-compacted latent representation is obtained by transforming, the more effective coding can be achieved. Unlike traditional codecs which use linear transformations, neural data compression is based on nonlinear transformations. Neural networks are capable of approximating arbitrary functions and can operate as a nonlinear transformation  \cite{leshno1993multilayer}. This property of neural networks provides the opportunity to transform the data with nonlinear dependency into a  more decorrelated representation.

Any improvement of the transformation function of a neural data compression algorithm can lead to coding supremacy. The transforming part of most neural data compression methods is based on convolutional neural networks, which have failed to take into account long-range dependencies. To address this shortage,  we propose to replace the convolutional network with a Transformer-based architecture. Our proposed Transformer framework leverages the self-attention module to capture global relationships. In addition, we equip our Transformer block with a Fused Local-aware Feed Forward (FLaFF) layer to strengthen the extraction of rich and diverse local textures, which is crucial for compression tasks. These enhancements can promote transformation's ability to project the data into a more decorrelated space.

\textbf{Contributions of this paper}. 
This paper presents a novel learned video compression approach specifically designed for compressing SDO images. The proposed algorithm aims to effectively exploit both spatial and temporal redundancies inherent in the dataset which enables the achievement of a high compression ratio. To enhance the capabilities of the non-linear transform and generate more decorrelated and energy-compacted latent code, we propose a Transformer-based transformation. Our proposed Transformer block leverages window-based self-attention modules and locally enhanced blocks, enabling the capture of both short-range and long-range relationships which are crucial for compression tasks. This approach also helps in reducing computational complexity. 

The remainder of the paper is organized as follows. Section II reviews the neural-based compression methods and the importance of compression in the SDO mission. Section III describes our proposed method. The experiments and ablation studies are discussed in section IV with a conclusion in section V.

%In this paper, we propose a novel learned video compression approach andapply it on SDO images. The proposed video compressionalgorithm aims to effectively leverage both spatial andtemporal redundancies inherent in the dataset, which enables the achievement of a high compression ratio. To enhance the capabilities of the non-linear transform in obtaining more decorrelated and energy-compacted latents, we extend the convolutional-based transformationto a Transformer-based architecture. The Transformerblock benefits from the window-based self-attention module and locallyenhanced block, enabling the capture of both short-rangeand long-range relationships, which is essential for compression tasks,   and reducing computational complexity .

\section{\textbf{Related Work}}\label{sec:related-work}

\begin{figure}[tp]
    \centering
    \includegraphics[width=0.75\linewidth]{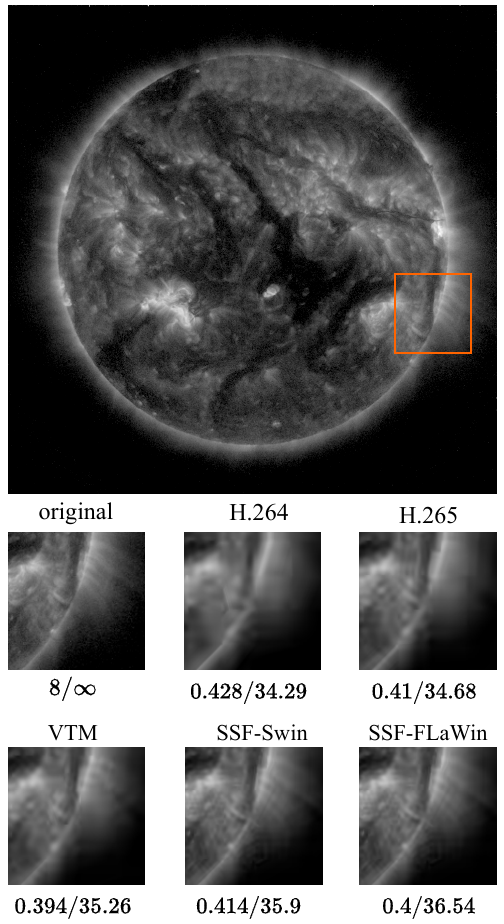}
    \caption{Visual comparison of the proposed neural video compression approaches (SSF-Swin and SSF-FLaWin) with other traditional codecs in terms of bit-rate/distortion [bpp$\downarrow$/PSNR$\uparrow$]. SSF-FLaWin demonstrates lower distortion in terms of PSNR compared to the other codecs, indicating its superior ability to preserve image quality. \emph{Best viewed on screen.}}
    \label{fig:visual-comparison}
\end{figure}

\subsection{\textbf{Neural Image Compression}}
Learned image compression methods often employ the transform coding scheme, which comprises four core steps \cite{goyal2001theoretical}. The first step utilizes an analysis transform to convert the input image into a compact and decorrelated latent representation. This transformation is crucial in reducing the data's redundancy. Once the latent representation is obtained, the second step involves quantization, where the continuous-valued latent variables are discretized to obtain discrete values. In the third step, entropy coding is employed, where an entropy model is utilized to generate a stream of ones and zeros. Finally, in the fourth step, a synthesis transform is applied to the quantized latent representations to reconstruct the original image \cite{balle2020nonlinear}. 
%By incorporating these four core steps, learned image compression methods can achieve high compression ratios while preserving image quality. 

%Initially, an analysis transform is utilized to transform the input image into a compact and decorrelated latent representation. Subsequently, the latent representation is quantized to obtain discrete values. The third step involves entropy coding, where an entropy model is employed to generate a stream of ones and zeros. Finally, a synthesis transform is applied to the quantized latent representations to reconstruct the original image [13].

Neural image compression networks commonly utilize the autoencoder architecture \cite{balle2020nonlinear}, which allows for the implementation of an approximately invertible nonlinear transformation. Alongside the transformation network, the entropy model is utilized for entropy coding, responsible for estimating the rate of the latent representation, and both are learned in an end-to-end fashion. However, learning the network parameters poses a challenge due to the non-differentiable nature of quantization, resulting in gradients that can be either zero or infinity. To address this issue, several solutions have been proposed to approximate quantization using differentiable operations \cite{balle2017endtoend,agustsson2017soft, yang2020variational}. A prevalent method to tackle the challenge of non-differentiable quantization is to replace it with additive uniform noise \cite{balle2017endtoend}. This substitution effectively transforms the autoencoder into a variational autoencoder (VAE) \cite{kingma2013auto} with a uniform encoder.

In early work, Ballé \emph{et al.} \cite{balle2016end} introduced the compressive autoencoder as a powerful image compression framework that achieved comparable performance to the JPEG2000 standard \cite{jpeg2000}. The compressive autoencoder employed the generalized divisive normalization (GDN) function to enable effective non-linear transformations and use a fully-factorized entropy model to accurately estimate the bit rate associated with the latent representation. To further improve the entropy model, Ballé et al. \cite{balle2018a} designed the hyperprior model, which conditions the distribution of the latent representation on hyperprior. This conditional distribution is approximated using a Gaussian scale mixture (GSM), with the scale parameters acquired from the decoding  hyperprior. Building upon this research, Minnen \emph{et al.} \cite{minnen2018joint} extended the entropy model from a Gaussian scale mixture to a Gaussian mixture model (GMM) by incorporating an autoregressive ingredient.

\subsection{\textbf{Neural Video Compression}}
Most neural video compression algorithms consist of two components: predictive coding and transform coding \cite{yang2022introduction}. Predictive coding is employed in inter-frame coding to exploit temporal redundancy. Inter-frame coding tries to predict the current frame from one or more previously reconstructed frames. In addition to inter-frame coding, intra-frame coding exists in the video compression pipeline, which leverages spatial redundancy to compress the frame. In intra-frame coding, the process  is analogous to image compression methods. As a result, frames are classified into three groups in video codecs: 1. I-frame (Intra-coded): compressed independently using image codecs; 2. P-frame (predicted): predicted from the past frames. 3. B-frames (bi-directional): predicted from the past and future frames \cite{rippel2019learned}.

\begin{figure*}[tp]
    \centering
    \includegraphics[width=0.82\linewidth]{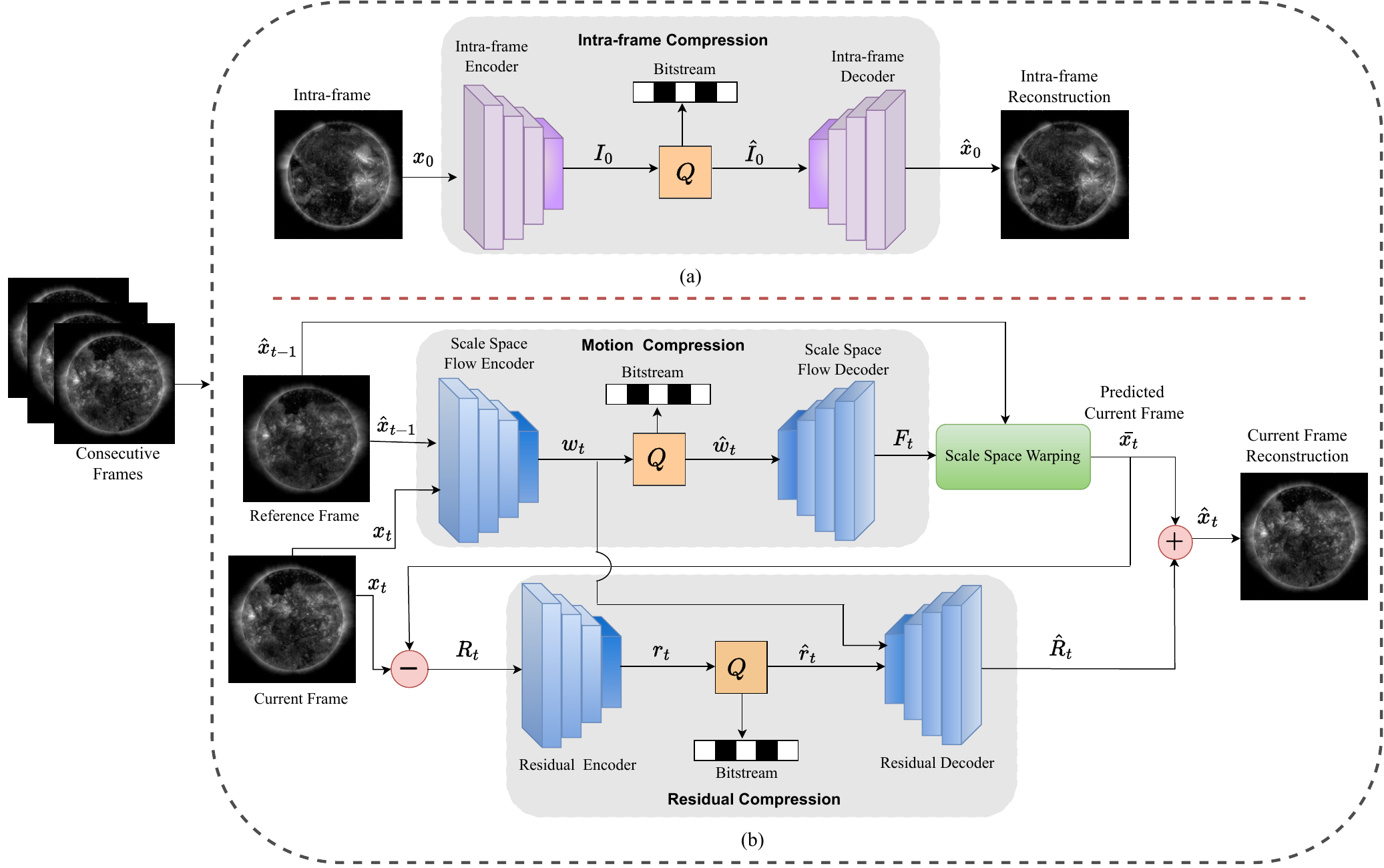}
    \caption{ An overview of neural video compression network. (a) The architecture of the I-frame compression model. (b)  The architecture of the P-frame compression model,  which consisting of motion compression and residual compression networks. The motion information and scale field are jointly estimated and encoded into a quantized latent representation $\hat{w}_{t}$. In the I-frame model, the previous reconstruction frame $\hat{x}_{t-1}$ is warped using the decoded motion and scale fields $F_t$, resulting in the prediction $\bar{x}_{t}$. The residual $R_t$ is then computed as the difference between the original current frame $x_t$ and the warped prediction $\bar{x}_{t}$. The residual is further encoded into a quantized latent representation $\hat{r}_{t}$, which is subsequently decoded to obtain $\hat{R}_{t}$. The final reconstructed current frame $\hat{x}_{t}$ is obtained by adding $\hat{R}_{t}$ to the warped prediction $\bar{x}_{t}$, resulting in $\hat{x}_{t} = \bar{x}_{t} + \hat{R}_{t}$. Entropy coding for each compression network is excluded for the sake of simplicity. }
    \label{fig:network-arch}
\end{figure*}

Recently, neural video compression methods have outperformed traditional video compression methods. Taking inspiration from the traditional hybrid video compression schemes, Lu \emph{et al.} \cite{lu2019dvc} introduced the Deep Video Compression (DVC) network as the first end-to-end deep video compression framework. This pioneering framework used a pre-trained Flownet \cite{ranjan2017optical} for optical flow estimation and employed bilinear warping techniques for motion compensation. For motion and residual compression, two autoencoder-based networks were used. 
Agustsson \emph{et al.} \cite{agustsson2020scale} proposed the Scale-Space Flow (SSF) framework, which aims to mitigate the difficulties associated with fast motion in optical flow estimation. Their approach involves the incorporation of a scale channel as an uncertainty parameter, allowing the application of Gaussian blur to regions prone to disocclusions and rapid motion.  Hu \emph{et al.} \cite{hu2021fvc} presented the Feature-space Video Compression (FVC) network as an advanced version of DVC. Their approach focuses on performing essential tasks, including motion estimation, motion compression, motion compensation, and residual compression, in the feature domain instead of the pixel space. In the framework presented in \cite{guo2021learning}, a cross-scale prediction module is incorporated to facilitate efficient motion compensation. Inspired by the observation that videos consist of a series of images with temporal redundancy, researchers \cite{habibian2019video, pessoa2020end} have extended image compression networks by adopting a 3D autoencoder-based framework to handle video data. The primary objective of this approach is to exploit spatial-temporal redundancies in videos by utilizing spatiotemporal transformations. To further enhance the performance of these networks, Habibian \emph{et al.} introduced a temporally conditional entropy model to leverage temporal correlations within the latent space.

In contrast to DVC, SSF, and FVC networks, which rely on a single previous frame as a reference frame, Lin \emph{et al.} \cite{lin2020m} propose a method that utilizes several previous frames to improve the accuracy of predicting the current frame. Mentzer \emph{et al.} \cite{mentzer2022neural} propose a neural video compression model based on Generative Adversarial Networks (GANs) \cite{goodfellow2020generative}. Their objective is to enhance the perceptual quality of the reconstructed frames by leveraging the power of GANs. More recently, Mentzer \emph{et al.} \cite{mentzer2022vct}present a novel network that avoids explicit motion estimation. Instead, they leverage a temporal transformer for entropy modeling.

\begin{figure*}[tp]
    \centering
    \includegraphics[width=0.81\linewidth]{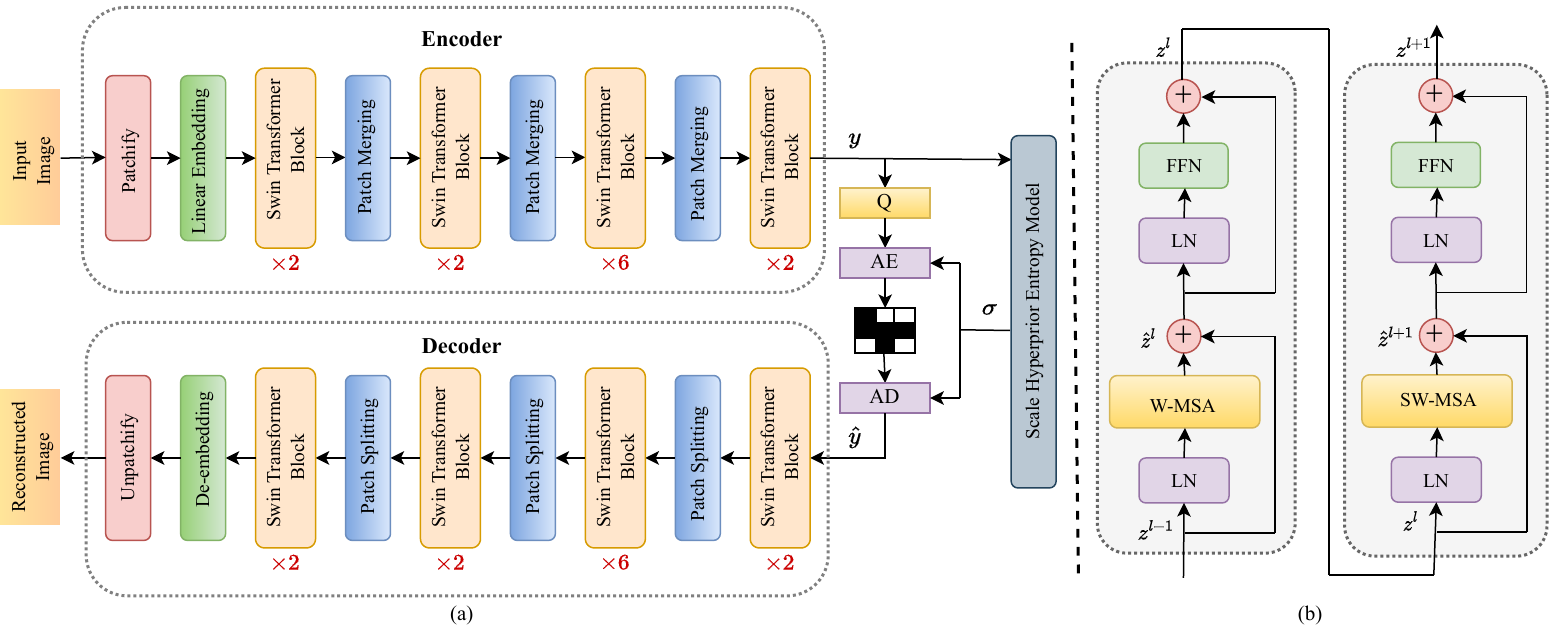}
    \caption{ Swin Transformer based architecture is designed for compressing I-frame, scale-space flow, and residual. (b) Two successive Swin Transformer blocks. Q shows scalar quantization. AE and AD refer to the arithmetic encoder and decoder, respectively.}
    \label{fig:network-arch}
\end{figure*}

\section{\textbf{Methods}}\label{sec:methods}

\subsection{\textbf{Overview}}
Our baseline model is the Scale-Space Flow (SSF) network \cite{agustsson2020scale}, which is one of the popular low-latency video compression models. As shown in Fig. 2, it is comprised of I-frame compression and P-frame compression models. The P-frame compression model consists of two parts: motion compression network and residual compression network. These three main networks, i.e., I-frame compression, motion compression, and residual compression are based on the autoencoder network architecture \cite{balle2017endtoend}.

The most important contribution of the SSF network is the generalization of optical flow to scale-space flow by adding a scale field to the motion field as a third channel. The scale field contributes to the model to blur the regions where disocclusion and fast motion exist, leading to a better inter-frame prediction. The inter-frame prediction is obtained by performing a trilinear warping on progressively blurred versions of the reference frame \cite{agustsson2020scale}. In the following sections, we will describe their novel components in P-frame compression.

\subsubsection{\textbf{Motion Compression}} The proposed motion compression network utilizes an autoencoder-based architecture, where the input consists of the current frame $\bm{x_t}$ and the previous reconstruction frame $\bm{\hat{x}_{t-1}}$. The encoder of the architecture is designed to jointly compute and encode the motion information that presents between the two consecutive frames. On the decoder side, the quantized latent representation of motion is decoded into three vectors: $F=(F_x, F_y, F_z)$.  The first two vectors, $F_x$ and $F_y$, represent horizontal and vertical motion vectors, respectively, and have dimensions of $\mathbb{R}^{2 \times H \times W}$. The third vector, $F_z$, corresponds to a one-channel scale field with dimensions of $\mathbb{R}^{H \times W}$.

\subsubsection{\textbf{Motion Compensation}} The motion compensation module plays a crucial role in predicting the current frame $\bm{\bar{x_t}}$ using the reference frame $\bm{\hat{x}_{t-1}}$ and the motion and scale fields. To achieve this, a scale-space warping operation is employed, where the reference frame undergoes progressive convolution with a Gaussian kernel. This convolution generates a series of blurred versions  of the reference frame:
\begin{equation}
   \bm{X}=[{\hat{{x}}_{t-1}},{\hat{{x}}_{t-1}}*G(s_1),...,{\hat{{x}}_{t-1}}*G(s_M)], 
\end{equation}
where $G(s_i)$ represents the Gaussian kernel with a scale parameter of $s_i$, the motion-compensated pixel value for each pixel located at the coordinate $[x, y]$ is obtained by applying trilinear interpolation in the scale-space volume. The process can be described as follows:
\begin{align}
\begin{split}
   &\bm{\bar{x_t}} =  \text{Scale-Space-Warp}(\hat{{x}}_{t-1},F)\\
   &{\bm{\bar{x_t}}[x,y]}=\bm{X}[x+F_{x}[x,y],y+F_{y}[x,y],F_{z}[x,y]].
\end{split}
\end{align}
\subsection{\textbf{Transformer-based Architecture}}

The Transformer \cite{vaswani2017attention} is originally proposed in the field of natural language processing (NLP) and had a profound impact on this domain. The remarkable accomplishments of the Transformer in NLP have motivated researchers to embrace the Transformer architecture in computer vision tasks. These tasks encompass a broad spectrum of applications, such as object detection \cite{carion2020end}, image classification \cite{touvron2021training}, semantic segmentation \cite{wang2021end}, and numerous other applications \cite{zafari2023frequency}. ViT  \cite{dosovitskiy2021an} is the first vision Transformer which utilizes a pure Transformer-based architecture for image classification and yields impressive results compared with traditional CNN networks \cite{he2016deep,simonyan2014very,sarlak2023diversity,BA2023}. It splits an image into non-overlapping patches and captures long-range dependencies by using multi-head self-attention module \cite{dosovitskiy2021an}. ViT has a high computational complexity due to the globally computed self-attention. The Swin Transformer \cite{liu2021swin} is proposed to reduce the computational complexity from quadratic to linear with respect to the patch numbers. The computational complexity is reduced because the Swin Transformer calculates self-attention locally within non-overlapping windows. The Swin Transformer network is also able to produce hierarchical representation which is very necessary for dense prediction tasks \cite{YuanFHLZCW21}. We have used the Swin Transformer to build the encoders and decoders of the Scale-Space Flow network. Fig. 3(a) shows the Swin Transformer-based architecture which is used to compress the I-frame, residual and scale-space flow. We have also extended the Swin Transformer block to the Fused Local-aware Window (FLaWin) Transformer block to enhance preserving local information.

%Transformer \cite{vaswani2017attention} is initially proposed in natural language processing (NLP) and has achieved promising performance in this domain. Thanks to the great success of transformer in the NLP field, researchers have adopted
%transformers architectur in various computer vision tasks such as image classification, object detection and semantic segmentation.

\subsection{\textbf{Swin Transformer-based Encoder/Decoder}}

%If the size of each patch is ${N\times N}$, the total number of patches becomes ${\frac{H}{N} \times \frac{W}{N}}$.

\subsubsection{\textbf{Encoder}}

The Swin Transformer, a pivotal component utilized as an encoder in the architecture \cite{liu2021swin}, comprises four essential blocks: Patchify, Linear Embedding, Swin Transformer block, and Patch merging. To initiate the encoding process, the input image ${x\in {R}^{C_{in}\times H \times W}}$ undergoes Patchify, which divides it into non-overlapping patches. Secondly, these patches are flattened and mapped into an embedding space with dimension ${C}$ by a linear embedding block. The output of these two blocks is then fed into the multiple Swin Transformer blocks and patch merging layers. The Swin Transformer block, a fundamental building block, is instrumental in maintaining the number of patches while efficiently extracting semantic features. This is achieved by performing local self-attention computations within each non-overlapping window, allowing the model to capture fine-grained meaningful information effectively. The patch merging layer generates hierarchical feature maps by halving the resolution of the feature map and doubling the channel number of the feature map, ensuring effective feature extraction and spatial information aggregation.

\begin{figure*}[tp]
    \centering
    \includegraphics[width=0.76\linewidth]{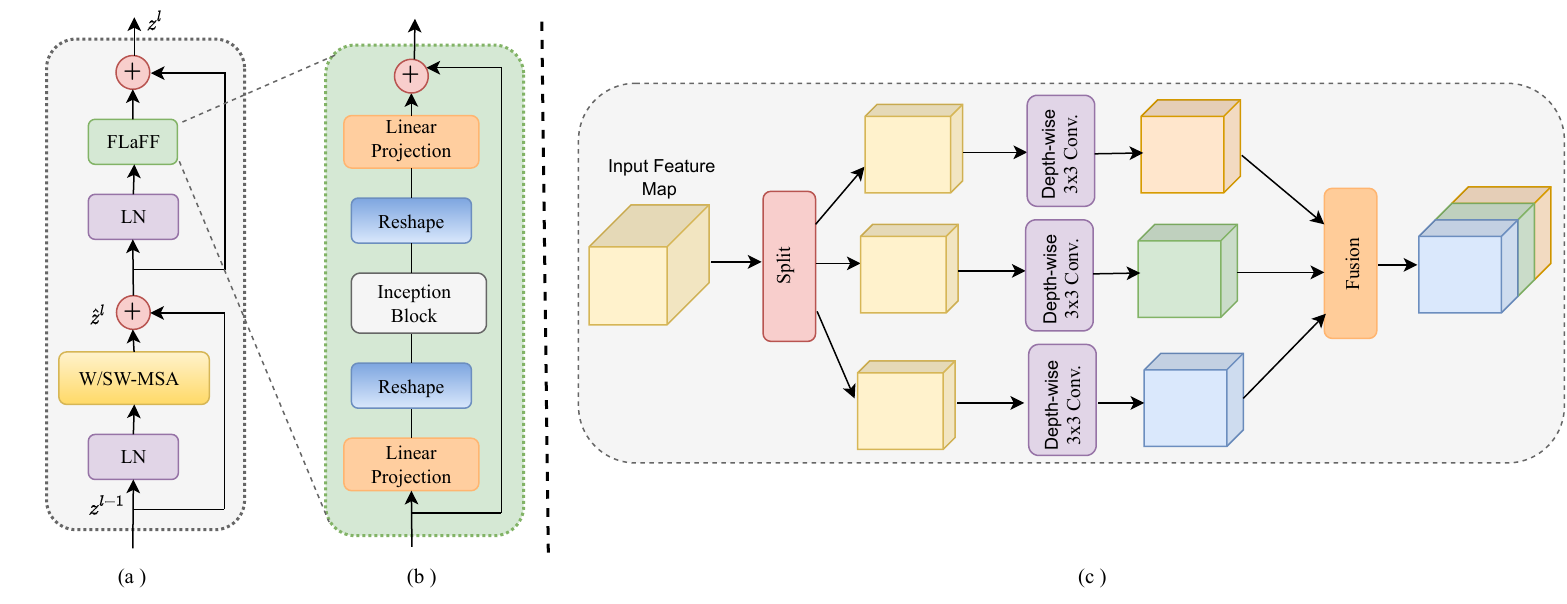}
    \caption{ (a) The architecture of FLaWin Transformer Block. (b) Fused local-aware feed-forward (FLaFF) network. (c) Structure of the proposed Inception block, which is used in FLaFF.}
    \label{fig:network-arch}
\end{figure*}

%The Swin Transformer \cite{liu2021swin}, which is utilized as an encoder, consists of four blocks: Patchify, Linear Embedding, Swin Transformer block, and Patch merging. First, the image ${x\in {R}^{ C_{in}\times H \times W}}$  is divided into non-overlapping patches through the patchify block.

%For example, patch merging, if its input feature map is ${x\in {R}^{ C \times \frac{H}{N} \times \frac{W}{N}}}$, groups each ${2\times 2}$ adjacent patches and concatenates them depth-wise, it leads to reduce the number of patches by factor of 4. Then, a linear layer is applyed on the ${4C}$-dimensional concatenated features to obtain ${2C}$-dimensional output.
%Swin transformer block is the main part of the swin transformer architecture. Unlike the traditional transformer block which consists of multi-head self-attention (MSA), swin transformer block is built based on the window based multi-head self-attention (W-MSA). Using W-MSA instead of MSA leads to a linear computational complexity since W-MSA conducts he self-attention within local windows. Although window-based multi-headed self-attention decreases the complexity, it has limitation in the information interaction across windows. To address this issue, shift-window-based multi-headed self-attention (SW-MSA) is employed after the W-MSA module. Therefore, the number of swin transformer blocks is always even, where W-MSA and SW-MSA are alternately applyed in successive  Swin transformer blocks.

\subsubsection{\textbf{Decoder}}
 The Swin Transformer decoder is the inverse symmetric of the encoder. We replace the patchify block with a unpatchify block, the patch merging layer with the patch splitting layer, and the linear embedding layer with a deembedding layer.
 \subsubsection{\textbf{Swin Transformer Block}}
The Swin Transformer block is the main part of the Swin Transformer architecture. Unlike the traditional Transformer block which is composed of multi-head self-attention (MSA), the Swin Transformer block is built upon a window-based multi-head self-attention (W-MSA) which conducts self-attention within local windows. The Window-based multi-head self-attention decreases the computational complexity; however, it fails to take into account the information interaction across different windows. To remedy this issue, the shifted-window-based multi-head self-attention (SW-MSA) is employed after the W-MSA module. As shown in Fig. 3(b), the Swin Transformer block consists of a layer normalization (LN), window-based multi-head self-attention (W-MSA) or shifted-window-based multi-head self-attention (SW-MSA), residual connection and a Feed-Forward Network (FFN), including a 2-layer MLP with GELU function as the nonlinearity. 
The process of two consecutive Transformer blocks can be defined as follows:
\begin{equation}
 \begin{aligned}
    &\bm{\hat z ^\mathnormal{l}} = {W\mbox{-}MSA(LN\bm(\bm{z ^{{\mathnormal{l}}-1}}}))+\bm{z ^{{\mathnormal{l}}-1}},
    \\& \bm{z ^\mathnormal l} = {MLP(LN(\bm{\hat z ^\mathnormal{l}}))}+ \bm{\hat z ^\mathnormal{l}},\\ &  \bm{\hat{z} ^{{\mathnormal{l}}+1}} = {SW\mbox{-}MSA(LN( \bm{z ^\mathnormal l}))}+\bm{z ^\mathnormal l},\\& \bm{{z} ^{{\mathnormal{l}}+1}} = {MLP(LN(\bm{\hat{z} ^{{\mathnormal{l}}+1}}))}+\bm{\hat{z} ^{{\mathnormal{l}}+1}},  
\end{aligned}
\end{equation}

where $\bm{\hat z ^\mathnormal{l}} $ and $\bm{\hat{z} ^{{\mathnormal{l}}+1}}$ show the outputs of W-MSA and SW-MSA  of the \emph{l}, and \emph{l+1} blocks, respectively. 
The self-attention mechanism employed in W-MAS and SW-MSA can be formulated as follows:
\begin{equation}
 \begin{aligned}
 Attention(\bm{Q},\bm{K}, \bm{V})= softmax( \frac{\bm{Q} \bm{{K}^T}}{\sqrt{d}}+\bm{B}) \bm{V},
 \end{aligned}
\end{equation}
Where $Q$, $K$, and $V\in {R}^{M^2 \times d
}$ show the query, key and value matrices respectively. The dimension of the key is denoted by $d$, and $M^2$ represents the number of patches in a window. The learnable relative position encoding is captured by the matrix $B$, which is derived from the bias matrix $B' \in \mathbb{R}^{(2M-1) \times (2M-1)}$ using learnable parameters. When there are $K$ attention heads, the attention mechanism is applied $K$ times in parallel, and the outputs of all heads are concatenated together. Finally, the concatenated outputs are linearly projected to obtain the final result.

%The process of Swin Transformer blocks is computed as:

%\begin{figure*}[tp]
    %\centering
    %\subfigure{\includegraphics[width=0.49\textwidth]{PSNR_NEW.%png}}
    %\hfill
   % \subfigure{\includegraphics[width=0.49\textwidth]{MS-ssimj.%png}}
    %\caption{Rate distortion curves on the test video clips. %Distortion is measured by PSNR (left) and MS-SSIM (right). %MS-SSIM metric is reported in logarithmic scale by $-10\log(1-m)$ to show the differences better, in which $m$ %is the MS-SSIM in the range of zero to one.}
    %\label{fig:rd-cvrves}
%\end{figure*}

\subsection{\textbf{Fused Local-aware Window Transformer Block}}
The feed-forward network (FFN) plays a crucial role in the Transformer block, known for its feature enhancement capabilities. In our proposed Transformer block, named Fused Local-aware Window (FLaWin), we replace the conventional FFN of the Swin Transformer block, comprising MLP layers, with our introduced fused local-aware feed forward (FLaFF) network. This incorporation of FLaFF in the Transformer block enables the capture of both local and long-range information while facilitating the extraction of diverse and multi-scale representations. Notably, the inclusion of local information is required for image compression tasks, where preserving fine-grained details is indispensable.

\subsubsection{\textbf{ Fused Local-aware Feed Forward (FLaFF)}} 

Our proposed FLaFF is composed of an Inception module which helps to extract local information and multi-scale representations. In the FLaFF architecture, as depicted in Fig. 4(b), first each token is passed through a linear projection layer, consisting of ${1\times 1}$ convolution layers, to increase its dimension. Second, the tokens are reshaped to a 2D token map, which is well-suited for the Inception block. Third, the Inception block is employed to extract diverse and local information from the 2-D token maps in parallel. Fourth, the 2D token maps are flattened and passed to another linear layer to project and lower the dimension of the input channels. 

%In summary, the aforementioned process can be expressed as follows:
%\begin{equation}
 %\begin{aligned}
   % &\bm{\hat{x}}=S2I(Conv_{1*1}(\bm{x})),
    %\\&\bm{\hat{x}}=Inception(\bm{\hat{x}}),\\ &\bm{x}= Conv_{1*1}%(I2S(\bm{\hat{x}}))+\bm{x}.
%\end{aligned}
%\end{equation}
%Where S2I represents the function which is responsible for reshaping the sequence of 1D tokens into a 2D feature map, and I2S performs the reverse process.

%In summary, the aforementioned process can be expressed as follows:

%\begin{equation}
 %\begin{aligned}
    %&\hat{x}=S2I(Conv_{1*1}(x)),
    %\\&\hat{x}=Inception-Block(\hat{x}),\\ &\hat{x}= %I2S(\hat{x}),\\
    %&x=Conv_{1*1}(\hat{x})+x.
%\end{aligned}
%\end{equation}

%Where S2I represents the function which is responsible for reshaping the sequence of 1D tokens into a 2D feature, and I2S performs the reverse process.

As illustrated in Fig. 4(c), the Inception block operates by dividing the 2-D input along the channel dimension and directing these split components into three separate branches. Each branch involves a depth-wise convolution with a kernel size of ${3\times 3}$. Utilizing depth-wise convolution in the Inception block offers two valuable benefits: it reduces computational complexity and enhances the modeling capabilities for channel attention. The convolution operations of these three branches are performed in parallel, and their outputs are concatenated along the channel dimension to form the final output of the Inception block. This architecture allows the FLaFF to effectively capture local details and diverse representations, significantly contributing to the overall performance.

\subsection{\textbf{Training Strategy}}

\subsubsection{\textbf{Loss Function}}
 The rate-distortion loss is used to train our network. If the length of the sequence is $T$, the total loss can be written as \cite{agustsson2020scale}:
\begin{equation}
D+\lambda R = \sum_{t=0}^{T-1} d({x_t},\hat{{x_t}})+ \lambda [R({I_0})+\sum_{t=1}^{T-1} R({w_t})+ R({r_t})],
\end{equation} where $D$ represents the distortion measure, such as the Mean Squared Error (MSE) between the original and reconstructed frames. $R$ denotes the bitrate to encode the quantized latent representation. ${I_0}$, ${w_t}$, ${r_t}$ represents I-frame, scale-space flow and residual latent, respectively. $\lambda$ is the Lagrangian coefficient that controls the trade-off between rate and distortion.
\subsubsection{\textbf{Quantization}}

To do entropy coding, quantization process need to be replaced with a soft differentiable function to make the end-to-end training feasible.
% Quantization is inevitable when it comes to entropy coding. Quantization is a non-differentiable and makes end-to-end training of the neural networks impossible; we are required to approximate the quantization function with a differentiable operation. 
In this paper, we add uniform noise to latent representations \cite{balle2017endtoend} to approximate the hard quantization during training. In the test phase, hard quantization i.e., a rounding operation is employed.
%\subsubsection{Entropy model}

\subsubsection{\textbf{Entropy Model}}
The entropy measurement should be used to estimate the bitrate for encoding the quantized latent representation. Therefore, it is required to estimate the probability distribution of quantized latent representation to compute the corresponding entropy. To do so, the hyper-prior network \cite{balle2018a} is utilized to estimate the probability distribution. The hyper-prior network proposes a hyper-prior latent representation ${z}$, as side information to capture the latent representation’s spatial dependencies. It results in computing the probability distribution of the latent representation precisely. The probability distribution of the quantized hyper-latent is estimated with a non-parametric fully factorized density model \cite{balle2018a}. The probability of quantized latent ${\hat{y}}$ conditioned on quantized hyper-prior ${\hat{z}}$ is modeled  by a zero-mean Gaussian distribution:
\begin{equation}
P_{\hat{y}|\hat{z}}(\hat{y}|\hat{z}){\sim} {\mathcal{N}}(0,\,\sigma^{2}),
\end{equation}
\noindent the scale parameter $\sigma$ is determined by the decoded quantized hyper-prior $\hat{z}$.

\section{Experiments}

%\subsection{Dataset} 
\subsection{\textbf{Dataset}} 
This research project relies on the extensive data collected during the Solar Dynamics Observatory (SDO) mission. The SDO mission is equipped with three instruments that operate continuously to capture essential information from the Sun \cite{sdoguide,nematirad2022solar,bhuveladesign,taghaviliquid}. The Helioseismic and Magnetic Imager (HMI) is specifically designed to study oscillations and the magnetic field present on the solar surface, known as the photosphere \cite{schou2012design}. It provides valuable insights into the dynamic behavior and magnetic properties of the Sun. The Atmospheric Imaging Assembly (AIA) captures full-sun images of the solar corona, covering a wide area of approximately 1.3 solar diameters. With a spatial resolution of around 1 arcsec, the AIA captures images at multiple wavelengths every 12 seconds \cite{lemen2012atmospheric}. This instrument offers detailed observations of the solar corona, which plays a significant role in understanding solar phenomena. To gain a deeper understanding of the variations that influence Earth's climate and near-Earth space, the Extreme Ultraviolet Variability Experiment (EVE) investigates the solar Extreme Ultraviolet (EUV) irradiance with high spectral precision \cite{woods2012extreme}. 

The original SDO dataset has undergone preprocessing  to generate a machine learning-ready dataset known as SDOML \cite{galvez2019machine}, which is used in this study. The SDOML dataset comprises AIA images captured at different wavelengths, including 94, 131, 171, 193, 211, 304, 335, 1600, and 1700 \angstrom\/ with a sampling rate of 6 minutes. In this paper, AIA images at the wavelength of 94  are utilized for both the training and testing phases. To train video compression networks on the SDOML dataset, we put four consecutive images together to make temporal chunks of four frames. Following the traditional codecs, during the test phase, we stack 30 consecutive images and create video clips with a GOP size of 30.

\begin{figure}
    
  \centering
  \scalebox{.37}{\includegraphics{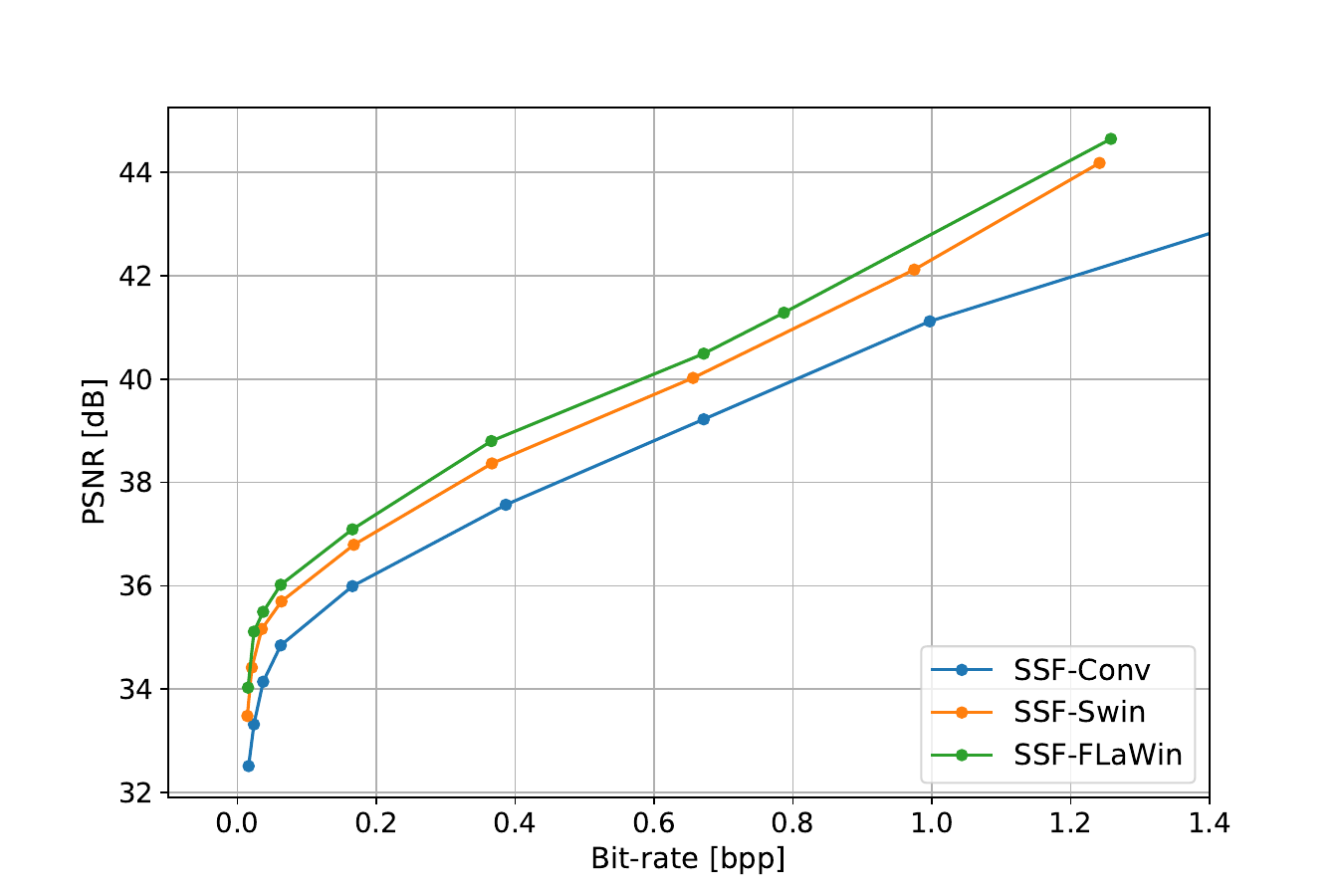}}
  \caption{The convolution-based model is compared with two different Transformer-based models, the Swin Transformer and FLaWin Transformer, in terms of the rate-distortion criteria.}
  \label{fig:qplot}
\end{figure}

%\subsection{Implementation Details} 
\subsection{\textbf{Implementation Details}} 

During the training process, our models are trained with a wide range of hyperparameters $\lambda\in\displaystyle\{0.00125, 0.0025, 0.005, 0.01, 0.02, 0.04, 0.08, 0.160, 0.320\}$ to cover various rate and distortion scenarios. The training is conducted for 100 epochs, with batches of size 16. Each batch comprises randomly cropped patches with dimensions of 256x256, extracted from the original 512x512 images. To optimize the model, we employ the Adam optimizer \cite{kingma2014adam} with an initial learning rate of $10^{-4}$, which gradually decreases to $1.2\times10^{-6}$ throughout the training process.

 \subsection{\textbf{Ablation Study}} 
To evaluate the effect of the Swin Transformer and how adding FLaFF to the Swin Transformer block help the video compression, we construct two versions of the Scale-Space Flow network. First, we replace the three convolutional autoencoders in I-frame and P-frame models with the Swin Transformer-based architecture. This model is called SSF-Swin. Then, we further enhance the Swin Transformer block by incorporating the proposed FLaWin Transformer block, which aims to improve the information extraction capability of the Transformer. This variant is termed SSF-FLaWin. As results are depicted in  Fig. 5, the SSF-Swin network achieves better performance in terms of rate-distortion trade-off compared to the SSF-Conv model. The superior performance of the SSF-Swin can be attributed to its ability to exploit the long-range correlations within the data, enabling better exploitation of spatial and temporal redundancies. Moreover, the introduction of the FLaWin Transformer block in SSF-FLaWin further contributes to the compression performance. The architectural design of FLaWin allows for the simultaneous capture of local details and long-range correlations, leading to the extraction of diverse and informative representations.

\begin{figure}
    
  \centering
  \scalebox{.37}{\includegraphics{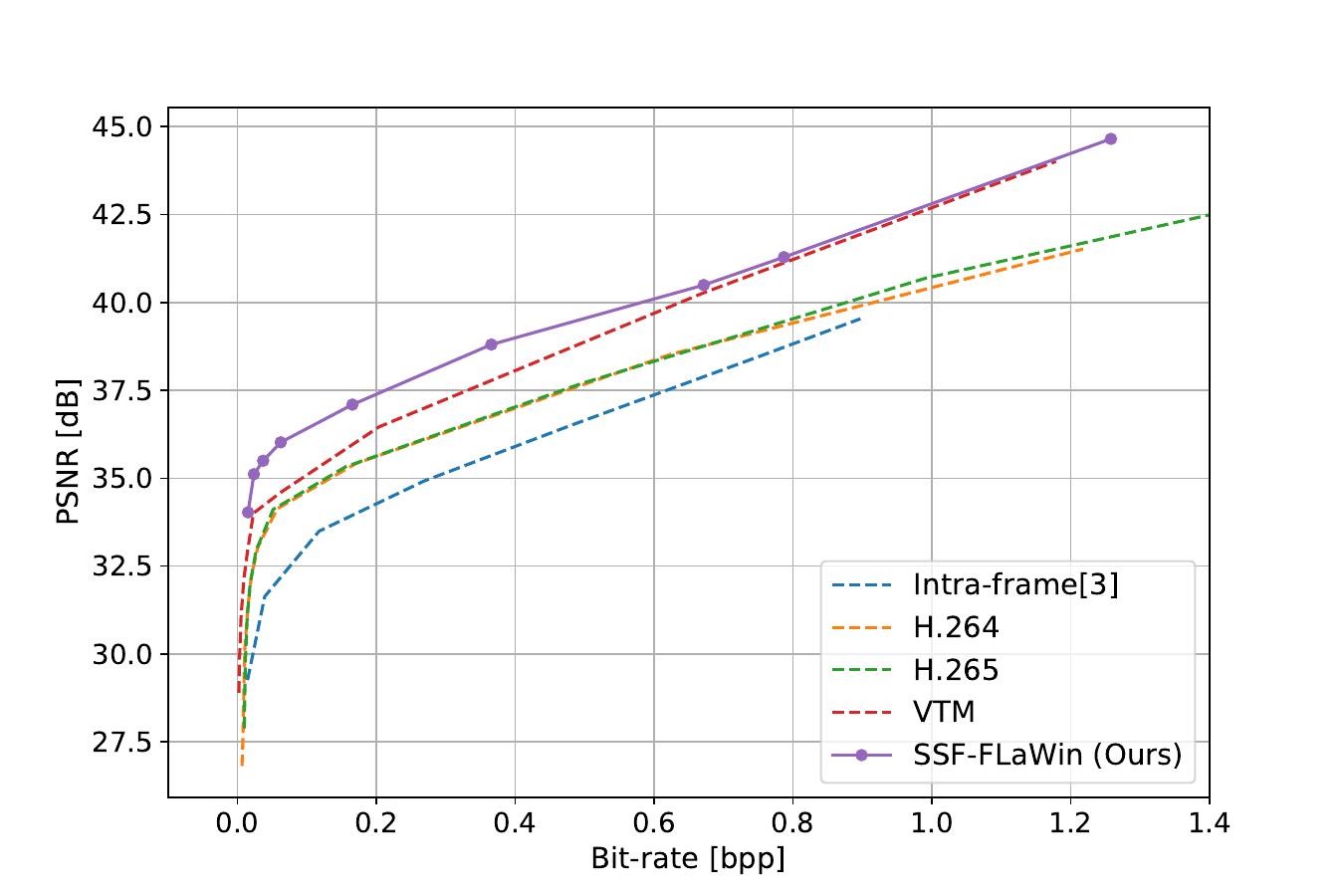}}
  \caption{ Rate distortion curves on the test video clips. Distortion is measured by PSNR.}
  \label{fig:qplot}
\end{figure}

 \subsection{\textbf{Comparison with the Traditional Video Codecs}}

 We conducted a comparison of the rate-distortion performance of our proposed network, SSF-FLaWin, with classical video compression standards and neural image compression on the SDOML dataset \cite{zafari2022attention,zafari2023att}, which serves as an equivalent to intra-frame compression. The distortion is measured by the Peak Signal-to-Noise Ratio (PSNR) metric. As depicted in Fig. 6, the rate-distortion performance of the neural video compression network, SSF-FLaWin, surpasses that of traditional video codecs such as H.264 and H.265, while achieving comparable performance with VTM \cite{vtm2022}. These findings clearly demonstrate the effectiveness of the FLaWin Transformer block in enhancing video compression performance. Furthermore, our results strongly emphasize the considerable advantage of video compression over image compression when applied to the SDOML dataset. This highlights the effectiveness of exploiting the temporal redundancies inherent in video data, which leads to significantly improved compression efficiency. 
 
%We compare the RD performance of our proposed network, SSF-FLaWin  with the classical video compression standards and neural image compression on SDOML dataset \cite{zafari2022attention}, which is equivalent to intra-frame compression. The distortion are measured by both the the Peak Signal-to-Noise Ratio (PSNR) metric Multi-scale structural similarity index Measure (MS-SSIM). As shown in Fig. 3, the RD performance of neural video compression network is better than he H.264 and H.265 and achieves comparable performance with VTM \cite{vtm2022}. These findings highlight the effectiveness of the LeWin Transformer block in enhancing video compression performance. Moreover, our findings strongly emphasize the considerable advantage of video compression over image compression when applied to the SDOML dataset. This highlights the effectiveness of exploiting the temporal redundancies inherent in video data, which leads to significantly improved compression efficiency.

\section{\textbf{Conclusion}}
 we have presented a Transformer-based neural video compression approach for the NASA'SDO mission. Our experimental results have clearly demonstrated the effectiveness of applying video compression techniques to the dataset, resulting in improved compression ratios. This improvement can be attributed to the high temporal correlation observed between the images in the dataset. Additionally, we have conducted an in-depth investigation into the coding efficiency of the Swin Transformer and FLaWin Transformer-based networks. The findings indicate the potential of these architectures for achieving efficient video compression. Overall, our work highlights the benefits of utilizing advanced Transformer models for enhancing video compression in the context of the SDO mission.

%\section*{\textbf{Acknowledgment}}
%This work was supported by NASA EPSCoR program.
\bibliographystyle{IEEEtran}
\bibliography{mybib}

\begin{thebibliography}{10}
\providecommand{\url}[1]{#1}
\csname url@samestyle\endcsname
\providecommand{\newblock}{\relax}
\providecommand{\bibinfo}[2]{#2}
\providecommand{\BIBentrySTDinterwordspacing}{\spaceskip=0pt\relax}
\providecommand{\BIBentryALTinterwordstretchfactor}{4}
\providecommand{\BIBentryALTinterwordspacing}{\spaceskip=\fontdimen2\font plus
\BIBentryALTinterwordstretchfactor\fontdimen3\font minus
  \fontdimen4\font\relax}
\providecommand{\BIBforeignlanguage}[2]{{%
\expandafter\ifx\csname l@#1\endcsname\relax
\typeout{** WARNING: IEEEtran.bst: No hyphenation pattern has been}%
\typeout{** loaded for the language `#1'. Using the pattern for}%
\typeout{** the default language instead.}%
\else
\language=\csname l@#1\endcsname
\fi
#2}}
\providecommand{\BIBdecl}{\relax}
\BIBdecl

\bibitem{schou2012hmi}
J.~Schou, P.~H. Scherrer, R.~I. Bush, R.~Wachter, S.~Couvidat, M.~C.
  Rabello-Soares, R.~S. Bogart, J.~Hoeksema, Y.~Liu, T.~Duvall \emph{et~al.},
  ``Design and ground calibration of the helioseismic and magnetic imager
  ({HMI}) instrument on the solar dynamics observatory ({SDO}),'' \emph{Solar
  Physics}, 2012.

\bibitem{fischer2017jpeg2000eve}
C.~E. Fischer, D.~M{\"u}ller, and I.~De~Moortel, ``{JPEG2000} image compression
  on solar {EUV} images,'' \emph{Solar Physics}, 2017.

\bibitem{zafari2022attention}
A.~Zafari, A.~Khoshkhahtinat, P.~M. Mehta, N.~M. Nasrabadi, B.~J. Thompson,
  D.~Da~Silva, and M.~S. Kirk, ``Attention-based generative neural image
  compression on solar dynamics observatory,'' in \emph{2022 21st IEEE
  International Conference on Machine Learning and Applications (ICMLA)}.\hskip
  1em plus 0.5em minus 0.4em\relax IEEE, 2022, pp. 198--205.

\bibitem{zafari2022neural}
A.~Zafari, A.~Khoshkhahtinat, N.~Nasrabadi, and P.~Mehta, ``Neural image
  compression on solar dynamics observatory,'' \emph{The Third Triennial
  Earth-Sun Summit (TESS}, vol.~54, no.~7, 2022.

\bibitem{yang2022introduction}
Y.~Yang, S.~Mandt, and L.~Theis, ``An introduction to neural data
  compression,'' \emph{CoRR}, 2022.

\bibitem{ma2019image}
S.~Ma, X.~Zhang, C.~Jia, Z.~Zhao, S.~Wang, and S.~Wang, ``Image and video
  compression with neural networks: A review,'' \emph{IEEE Transactions on
  Circuits and Systems for Video Technology}, 2019.

\bibitem{leshno1993multilayer}
M.~Leshno, V.~Y. Lin, A.~Pinkus, and S.~Schocken, ``Multilayer feedforward
  networks with a nonpolynomial activation function can approximate any
  function,'' \emph{Neural networks}, 1993.

\bibitem{goyal2001theoretical}
V.~K. Goyal, ``Theoretical foundations of transform coding,'' \emph{IEEE Signal
  Processing Magazine}, vol.~18, no.~5, pp. 9--21, 2001.

\bibitem{balle2020nonlinear}
J.~Ball{\'e}, P.~A. Chou, D.~Minnen, S.~Singh, N.~Johnston, E.~Agustsson, S.~J.
  Hwang, and G.~Toderici, ``Nonlinear transform coding,'' \emph{IEEE Journal of
  Selected Topics in Signal Processing}, vol.~15, no.~2, pp. 339--353, 2020.

\bibitem{balle2017endtoend}
J.~Ball{\'{e}}, V.~Laparra, and E.~P. Simoncelli, ``End-to-end optimized image
  compression,'' in \emph{ICLR}, 2017.

\bibitem{agustsson2017soft}
E.~Agustsson, F.~Mentzer, M.~Tschannen, L.~Cavigelli, R.~Timofte, L.~Benini,
  and L.~V. Gool, ``Soft-to-hard vector quantization for end-to-end learning
  compressible representations,'' \emph{Advances in neural information
  processing systems}, vol.~30, 2017.

\bibitem{yang2020variational}
Y.~Yang, R.~Bamler, and S.~Mandt, ``Variational bayesian quantization,'' in
  \emph{International Conference on Machine Learning}.\hskip 1em plus 0.5em
  minus 0.4em\relax PMLR, 2020, pp. 10\,670--10\,680.

\bibitem{kingma2013auto}
D.~P. Kingma and M.~Welling, ``Auto-encoding variational bayes,'' \emph{arXiv
  preprint arXiv:1312.6114}, 2013.

\bibitem{balle2016end}
J.~Ball{\'e}, V.~Laparra, and E.~P. Simoncelli, ``End-to-end optimized image
  compression,'' \emph{arXiv preprint arXiv:1611.01704}, 2016.

\bibitem{jpeg2000}
D.~S. Taubman and M.~W. Marcellin, \emph{{JPEG2000} - image compression
  fundamentals, standards and practice}, ser. The Kluwer international series
  in engineering and computer science.\hskip 1em plus 0.5em minus 0.4em\relax
  Kluwer, 2002.

\bibitem{balle2018a}
J.~Ball{\'{e}}, D.~Minnen, S.~Singh, S.~J. Hwang, and N.~Johnston,
  ``Variational image compression with a scale hyperprior,'' in \emph{ICLR},
  2018.

\bibitem{minnen2018joint}
D.~Minnen, J.~Ball{\'e}, and G.~D. Toderici, ``Joint autoregressive and
  hierarchical priors for learned image compression,'' \emph{Advances in neural
  information processing systems}, vol.~31, 2018.

\bibitem{rippel2019learned}
O.~Rippel, S.~Nair, C.~Lew, S.~Branson, A.~G. Anderson, and L.~Bourdev,
  ``Learned video compression,'' in \emph{ICCV}, 2019.

\bibitem{lu2019dvc}
G.~Lu, W.~Ouyang, D.~Xu, X.~Zhang, C.~Cai, and Z.~Gao, ``{DVC}: An end-to-end
  deep video compression framework,'' in \emph{Proceedings of the IEEE/CVF
  Conference on Computer Vision and Pattern Recognition}, 2019, pp.
  11\,006--11\,015.

\bibitem{ranjan2017optical}
A.~Ranjan and M.~J. Black, ``Optical flow estimation using a spatial pyramid
  network,'' in \emph{Proceedings of the IEEE conference on computer vision and
  pattern recognition}, 2017, pp. 4161--4170.

\bibitem{agustsson2020scale}
E.~Agustsson, D.~Minnen, N.~Johnston, J.~Balle, S.~J. Hwang, and G.~Toderici,
  ``Scale-space flow for end-to-end optimized video compression,'' in
  \emph{Proceedings of the IEEE/CVF Conference on Computer Vision and Pattern
  Recognition}, 2020, pp. 8503--8512.

\bibitem{hu2021fvc}
Z.~Hu, G.~Lu, and D.~Xu, ``{FVC}: A new framework towards deep video
  compression in feature space,'' in \emph{Proceedings of the IEEE/CVF
  Conference on Computer Vision and Pattern Recognition}, 2021, pp. 1502--1511.

\bibitem{guo2021learning}
Z.~Guo, R.~Feng, Z.~Zhang, X.~Jin, and Z.~Chen, ``Learning cross-scale
  prediction for efficient neural video compression,'' \emph{arXiv e-prints},
  2021.

\bibitem{habibian2019video}
A.~Habibian, T.~v. Rozendaal, J.~M. Tomczak, and T.~S. Cohen, ``Video
  compression with rate-distortion autoencoders,'' in \emph{Proceedings of the
  IEEE/CVF International Conference on Computer Vision}, 2019, pp. 7033--7042.

\bibitem{pessoa2020end}
J.~Pessoa, H.~Aidos, P.~Tom{\'a}s, and M.~A. Figueiredo, ``End-to-end learning
  of video compression using spatio-temporal autoencoders,'' in \emph{2020 IEEE
  Workshop on Signal Processing Systems (SiPS)}.\hskip 1em plus 0.5em minus
  0.4em\relax IEEE, 2020, pp. 1--6.

\bibitem{lin2020m}
J.~Lin, D.~Liu, H.~Li, and F.~Wu, ``{M-LVC}: Multiple frames prediction for
  learned video compression,'' in \emph{Proceedings of the IEEE/CVF Conference
  on Computer Vision and Pattern Recognition}, 2020, pp. 3546--3554.

\bibitem{mentzer2022neural}
F.~Mentzer, E.~Agustsson, J.~Ball{\'e}, D.~Minnen, N.~Johnston, and
  G.~Toderici, ``Neural video compression using gans for detail synthesis and
  propagation,'' in \emph{European Conference on Computer Vision}.\hskip 1em
  plus 0.5em minus 0.4em\relax Springer, 2022, pp. 562--578.

\bibitem{goodfellow2020generative}
I.~Goodfellow, J.~Pouget-Abadie, M.~Mirza, B.~Xu, D.~Warde-Farley, S.~Ozair,
  A.~Courville, and Y.~Bengio, ``Generative adversarial networks,''
  \emph{Communications of the ACM}, vol.~63, no.~11, pp. 139--144, 2020.

\bibitem{mentzer2022vct}
F.~Mentzer, G.~Toderici, D.~Minnen, S.-J. Hwang, S.~Caelles, M.~Lucic, and
  E.~Agustsson, ``{VCT}: A video compression transformer,'' \emph{arXiv
  preprint arXiv:2206.07307}, 2022.

\bibitem{vaswani2017attention}
A.~Vaswani, N.~Shazeer, N.~Parmar, J.~Uszkoreit, L.~Jones, A.~N. Gomez,
  {\L}.~Kaiser, and I.~Polosukhin, ``Attention is all you need,''
  \emph{Advances in neural information processing systems}, vol.~30, 2017.

\bibitem{carion2020end}
N.~Carion, F.~Massa, G.~Synnaeve, N.~Usunier, A.~Kirillov, and S.~Zagoruyko,
  ``End-to-end object detection with transformers,'' in \emph{European
  conference on computer vision}.\hskip 1em plus 0.5em minus 0.4em\relax
  Springer, 2020, pp. 213--229.

\bibitem{touvron2021training}
H.~Touvron, M.~Cord, M.~Douze, F.~Massa, A.~Sablayrolles, and H.~J{\'e}gou,
  ``Training data-efficient image transformers \& distillation through
  attention,'' in \emph{International conference on machine learning}.\hskip
  1em plus 0.5em minus 0.4em\relax PMLR, 2021, pp. 10\,347--10\,357.

\bibitem{wang2021end}
Y.~Wang, Z.~Xu, X.~Wang, C.~Shen, B.~Cheng, H.~Shen, and H.~Xia, ``End-to-end
  video instance segmentation with transformers,'' in \emph{Proceedings of the
  IEEE/CVF conference on computer vision and pattern recognition}, 2021, pp.
  8741--8750.

\bibitem{zafari2023frequency}
A.~Zafari, A.~Khoshkhahtinat, P.~Mehta, M.~S.~E. Saadabadi, M.~Akyash, and
  N.~M. Nasrabadi, ``Frequency disentangled features in neural image
  compression,'' in \emph{2023 IEEE International Conference on Image
  Processing (ICIP)}.\hskip 1em plus 0.5em minus 0.4em\relax IEEE, 2023, pp.
  2815--2819.

\bibitem{dosovitskiy2021an}
A.~Dosovitskiy, L.~Beyer, A.~Kolesnikov, D.~Weissenborn, X.~Zhai,
  T.~Unterthiner, M.~Dehghani, M.~Minderer, G.~Heigold, S.~Gelly, J.~Uszkoreit,
  and N.~Houlsby, ``An image is worth 16x16 words: Transformers for image
  recognition at scale,'' in \emph{ICLR}, 2021.

\bibitem{he2016deep}
K.~He, X.~Zhang, S.~Ren, and J.~Sun, ``Deep residual learning for image
  recognition,'' in \emph{Proceedings of the IEEE conference on computer vision
  and pattern recognition}, 2016, pp. 770--778.

\bibitem{simonyan2014very}
K.~Simonyan and A.~Zisserman, ``Very deep convolutional networks for
  large-scale image recognition,'' \emph{arXiv preprint arXiv:1409.1556}, 2014.

\bibitem{sarlak2023diversity}
A.~Sarlak, A.~Razi, X.~Chen, and R.~Amin, ``Diversity maximized scheduling in
  roadside units for traffic monitoring applications,'' in \emph{2023 IEEE 48th
  Conference on Local Computer Networks (LCN)}.\hskip 1em plus 0.5em minus
  0.4em\relax IEEE, 2023, pp. 1--4.

\bibitem{BA2023}
B.~Adami, S.~Tehranipoor, N.~M. Nasrabadi, and N.~Karimian, ``A universal
  anti-spoofing approach for contactless fingerprint biometric systems,'' in
  \emph{2023 IEEE International Joint Conference on Biometrics (IJCB)}.\hskip
  1em plus 0.5em minus 0.4em\relax IEEE, 2023, pp. 1--8.

\bibitem{liu2021swin}
Z.~Liu, Y.~Lin, Y.~Cao, H.~Hu, Y.~Wei, Z.~Zhang, S.~Lin, and B.~Guo, ``Swin
  transformer: Hierarchical vision transformer using shifted windows,'' in
  \emph{ICCV}, 2021.

\bibitem{YuanFHLZCW21}
Y.~Yuan, R.~Fu, L.~Huang, W.~Lin, C.~Zhang, X.~Chen, and J.~Wang,
  ``{HRFormer}:high-resolution transformer for dense prediction,''
  \emph{NeurIPS}, 2021.

\bibitem{sdoguide}
\BIBentryALTinterwordspacing
``A guide to the mission and purpose of nasa’s solar dynamics observatory,''
  2010. [Online]. Available:
  \url{{https://sdo.gsfc.nasa.gov/assets/docs/SDO\_Guide.pdf}}
\BIBentrySTDinterwordspacing

\bibitem{nematirad2022solar}
R.~Nematirad and A.~Pahwa, ``Solar radiation forecasting using artificial
  neural networks considering feature selection,'' in \emph{2022 IEEE Kansas
  Power and Energy Conference (KPEC)}.\hskip 1em plus 0.5em minus 0.4em\relax
  IEEE, 2022, pp. 1--4.

\bibitem{bhuveladesign}
P.~Bhuvela and A.~Nasiri, ``Design methodology for a medium voltage single
  stage llc resonant solar pv inverter.''

\bibitem{taghaviliquid}
H.~Taghavi, A.~El~Shafei, and A.~Nasiri, ``Liquid cooling system for a high
  power, medium frequency, and medium voltage isolated power converter.''

\bibitem{schou2012design}
J.~Schou, P.~H. Scherrer, R.~I. Bush, R.~Wachter, S.~Couvidat, M.~C.
  Rabello-Soares, R.~Bogart, J.~Hoeksema, Y.~Liu, T.~Duvall \emph{et~al.},
  ``Design and ground calibration of the {Helioseismic and Magnetic Imager
  (HMI) instrument on the Solar Dynamics Observatory (SDO)},'' \emph{Solar
  Physics}, vol. 275, pp. 229--259, 2012.

\bibitem{lemen2012atmospheric}
J.~R. Lemen, A.~M. Title, D.~J. Akin, P.~F. Boerner, C.~Chou, J.~F. Drake,
  D.~W. Duncan, C.~G. Edwards, F.~M. Friedlaender, G.~F. Heyman \emph{et~al.},
  ``The atmospheric imaging assembly {(AIA)} on the solar dynamics observatory
  ({SDO}),'' \emph{Solar Physics}, vol. 275, pp. 17--40, 2012.

\bibitem{woods2012extreme}
T.~N. Woods, F.~Eparvier, R.~Hock, A.~Jones, D.~Woodraska, D.~Judge,
  L.~Didkovsky, J.~Lean, J.~Mariska, H.~Warren \emph{et~al.}, ``{Extreme
  Ultraviolet Variability Experiment (EVE) on the Solar Dynamics Observatory
  (SDO)}: Overview of science objectives, instrument design, data products, and
  model developments,'' \emph{The solar dynamics observatory}, pp. 115--143,
  2012.

\bibitem{galvez2019machine}
R.~Galvez, D.~F. Fouhey, M.~Jin, A.~Szenicer, A.~Mu{\~n}oz-Jaramillo, M.~C.
  Cheung, P.~J. Wright, M.~G. Bobra, Y.~Liu, J.~Mason \emph{et~al.}, ``A
  machine-learning data set prepared from the {NASA} solar dynamics observatory
  mission,'' \emph{The Astrophysical Journal Supplement Series}, 2019.

\bibitem{kingma2014adam}
D.~P. Kingma and J.~Ba, ``Adam: A method for stochastic optimization,''
  \emph{arXiv preprint arXiv:1412.6980}, 2014.

\bibitem{zafari2023att}
A.~Zafari, A.~Khoshkhahtinat, P.~M. Mehta, N.~Nasrabadi, B.~J. Thompson,
  D.~da~Silva, and M.~Kirk, ``Attention-based generative neural image
  compression on solar dynamics observatory,'' in \emph{103rd AMS Annual
  Meeting}.\hskip 1em plus 0.5em minus 0.4em\relax AMS, 2023.

\bibitem{vtm2022}
``{Versatile Video Coding Reference Software},'' Available at
  \url{https://vcgit.hhi.fraunhofer.de/jvet/VVCSoftware_VTM}, 2022.

\end{thebibliography}
\end{document}